\documentstyle[prl,aps,epsfig]{revtex}

\begin{document}
\twocolumn[\hsize\textwidth\columnwidth\hsize\csname
@twocolumnfalse\endcsname \draft
\author{E. Solano,$^{1,3}$  R. L. de Matos Filho,$^{2}$ and N. Zagury
$^{2}$}
\title{Entangled coherent states and squeezing in $N$ trapped ions}
\address{$^{1}$Max-Planck-Institut f{\"u}r Quantenoptik,
Hans-Kopfermann-Strasse 1, 85748 Garching, Germany
\\ $^{2}$Instituto de F\'{\i}sica, Universidade Federal do Rio de
Janeiro,
Caixa Postal 68528, 21945-970 Rio de Janeiro, RJ, Brazil \\
$^{3}$Secci\'{o}n F\'{\i}sica, Departamento de Ciencias,
Pontificia Universidad Cat\'{o}lica del Per\'{u}, Apartado 1761,
Lima, Peru}
\date{\today}
\maketitle

\begin{abstract}
We consider a resonant bichromatic excitation of $N$ trapped ions
that generates displacement and squeezing in their collective
motion conditioned to their ionic internal state, producing
eventually Schr\"odinger cat states and entangled squeezing.
Furthermore, we study the case of tetrachromatic illumination for
producing the so called entangled coherent states in two motional
normal modes.
\end{abstract}

]

\vskip2pc

In the last years great attention has being given to the
possibility of producing mesoscopic superposition of states where
a great number of particles and degrees of freedom is
involved~\cite{ioncats,CQEDcats,fourionent,threeatoment}. Beyond
their possible applications in quantum
information~\cite{reviewions,reviewCQED}, they represent an
important tool in the experimental studies of decoherence and the
emergence of the classical world out of its quantized
version~\cite{Zurek}. In the microscopic world, a system can exist
in a superposition of different quantum states, given rise to
interference effects. However, these superpositions do not
manifest themselves in the classical world~\cite{Schrodinger}.
This paradox is partially solved by realizing that coherence is
lost increasingly fast with the size of the
system~\cite{CQEDcats,decoherenceions}. Therefore, the generation
of mesoscopic quantum superpositions, and the study of the time
scale in which decoherence occurs, is an important step for
understanding the boundary between classical physics and quantum
mechanics. The system of $N$ trapped ions is a good laboratory for
building large quantum superpositions and analyzing how
decoherence appears when the system grows, since it is weakly
affected by the environment~\cite{reviewions,decoherenceions}. In
this work, we discuss some proposals for producing, in a fast and
controllable way, vibronic mesoscopic superpositions and other
nonclassical states in a system consisting of $N$ trapped
ions~\cite{cation,santiagoreview}. In particular, generation of
Schr\"odinger cat states, entangled squeezing and entangled
coherent states are discussed.

We consider the situation where $N$ two-level ions of mass $m$ are
confined to move in the $z$ direction of a Paul trap. We assume
that they are cooled down to low temperatures~\cite
{fourionent,king} and able to perform small oscillations around
their equilibrium positions, $z_{j0}$ ($ j=1,2...N$). Two
classical fields, $\vec{E}_{I}= \vec{E}_{0I}e^{i(q_{I} z -\omega
_{I}t-\varphi_I)}$ and $\vec{E}_{II} = \vec{E}_{0II}\,e^{i(q_{II}
z-\omega _{II}t-\varphi_{II})}$, illuminate homogeneously all $N$
ions, with angular frequencies $\omega_{I}$ and $\omega_{II}$,
wave vectors, $q_{I}=q_{II}=q$, parallel to the $z$ direction, and
phases $\varphi_I=\varphi_{II}=\varphi$. Depending on the
experimental setup, each of theses fields may be obtained
effectively from a pair of lasers in Raman configuration. The
angular frequencies $\omega_{I}$ and $\omega_{II}$ are chosen to
be quasiresonant with a ionic internal transition between
long-living levels $|e_{j} \rangle $ and $|g_{j} \rangle $
$(j=1,...N),$ with energies $\hbar\omega_0$ and $0$, respectively.
The total Hamiltonian of the system may be written, in the optical
rotating wave approximation (RWA), as
\begin{eqnarray}
\hat{H}=\hat{H}_0+\hat{H}_{{\rm int}},
\end{eqnarray}
with
\begin{eqnarray}\label{H0}
\hat{H}_0=\hbar \omega_0\sum_{j=1,N}|e_j \rangle \langle e_j| +
\hbar \nu a^\dagger a + \sum_{\lambda=1,N-1} \hbar\nu_\lambda
b_\lambda^\dagger b_\lambda\, \nonumber ,
\end{eqnarray}
and
\begin{eqnarray}\label{Hint}
\hat{H}_{\rm int}=\hbar \Omega \sum_{j=1,N} && \left[
e^{i(q\hat{z}_j -i\omega_{I}t - \varphi)} + e^{i(q\hat{z}_j
-i\omega_{II}t - \varphi)} \right] | e_j\rangle\langle g_j| \nonumber \\
&& + {\rm H.c.}
\end{eqnarray}
The operators $\{ a , b_\lambda \}$ and $\{ a^{\dagger} ,
b_\lambda^{\dagger } \}$ are the annihilation and creation
operators associated with the center-of-mass (CM) mode of
frequency $\nu$ and with the $N-1$ other vibrational modes of
frequency $\nu_\lambda$, respectively~\cite{James}. We assumed
also that the same real coupling constant $\Omega$ appears in the
interaction of each excitation field with each ion.

We choose the  frequencies $\omega_I$ and $\omega_{II}$ to be
\begin{eqnarray}  \label{freq}
\omega_{I} = \omega_0 + k\nu  \quad {\rm and} \quad \omega_{II}
=\omega_0 -k\nu\, ,
\end{eqnarray}
so that we excite the CM vibronic transition in the $k$-th blue
and $k$-th red motional
sideband~\cite{sashavogel,moyacessasashavogel}. We go to the
interaction picture and make a RWA with respect to the CM
vibrational frequency, selecting the terms that oscillate with
minimum frequency~\cite{vogelruynet}. For simplifying the
notation, we set $\varphi = k \pi /2$ and define the "spin angular
momentum states" $|\uparrow_j\rangle$$=e^{iqz_{j0}} |e_j \rangle$
and $|\downarrow_j\rangle=|g_j\rangle$. Then, in the Lamb-Dicke
regime, the resulting effective Hamiltonian may be written as
\begin{eqnarray}
\label{geral} H_{\rm int} = \frac{\hbar \eta^k \Omega}{k!}
\sum_{j=1}^N ( {\sigma_{j}}^{\dagger} + \sigma_{j} ) (a^k +
a^{\dagger k }) ,
\end{eqnarray}
where $\eta = q \sqrt{\hbar /2Nm\nu }$ is the CM Lamb-Dicke
parameter and $\widehat{\sigma}_{+j}=| \! \uparrow_j \rangle
\langle \downarrow_j \! |$ $=e^{iqz_{j0}} | e_j\rangle\langle g_j
|$ is the so redefined flip-up operator. Note that, in the
Hamiltonian of Eq.~(\ref{geral}), the operators associated with
the internal and external degrees of freedom are decoupled. In
fact, it could be rewritten as
\begin{eqnarray}
\label{geralangular} H_{\rm int} = \frac{2 \hbar \eta^k
\Omega}{k!} \,\, {J}_x (a^k + a^{\dagger k }) ,
\end{eqnarray}
where $J_x = \sum_{j=1}^N ( {\sigma_{j}}^{\dagger} + \sigma_{j} )
/ 2$ may be associated with the $x$ component of an angular
momentum operator $\vec J$, being the $y$ and $z$ components $J_y
= \sum_{j=1}^N ( \sigma_{j}^{\dagger}-{\sigma_{j}})/ (2i)$ and
$J_z = \sum_{j=1}^N (|e_j\rangle\langle e_j|-|g_j\rangle\langle
g_j|)/2$, respectively. We now express the evolution operator, at
time $t$, as a sum of products of unitary operators acting on the
motional state and projection operators acting on the internal
state of the ions
\begin{eqnarray}\label{evoltime}
U_k(t)=\sum_{j,m } D_k \lbrack m \, \Omega_k t \rbrack \,\,
|{j,m}\rangle_{x\,\, x} \langle{j,m }| ,
\end{eqnarray}
with $\Omega_k = 2 i \eta^k \Omega /k!$. Also,
\begin{equation}\label{displacement}
{D}_k(\chi) = e^{\chi a^{\dagger k} - \chi^* a^k}
\end{equation}
and $|j,m\rangle_x$ are the simultaneous eigenvectors of the
operators $J_x$ and $J^2\equiv J_x^2 + J_y^2 + J_z^2$ associated
with the eigenvalues $m=-j,-(j-1), .....j$ and $j(j+1)$,
respectively. $j$ varies from $0$  $(1/2)$ to $N/2$ by steps of
$1,$ if $N$ is even (odd).

Eq.~(\ref{evoltime}) shows an important feature of the proposed
scheme, it yields an evolution that corresponds to unitary
operations $\hat{D}_k(\chi)$ on the CM mode conditioned to the
value $m$ of the $x$-component of the "angular momentum"
electronic state. When $k=1$, the excitations occuring in the
first red and first blue sidebands, Eq.~(\ref{displacement}) turns
into the familiar displacement operator $D(\alpha)$, which
produces the coherent state $| \alpha \rangle$ out of an initial
vacuum state $| 0 \rangle$. For example, let us consider the case
$N = 1$ with an initial state $| \downarrow \rangle | 0 \rangle =
( | + \rangle + | - \rangle ) / \sqrt{2}$, where the sates $| \pm
\rangle = ( | \downarrow \rangle \pm | \uparrow \rangle ) /
\sqrt{2}$ are the eigenstates of the operator $J_x$ for the case
of a single spin $1 / 2$. After a time $\tau$, the evolved state
will be
\begin{eqnarray}
\label{cats} \frac{1}{\sqrt{2}} \big( | + \rangle | \alpha \rangle
+ | - \rangle | - \alpha \rangle \big) ,
\end{eqnarray}
where $\alpha= i \eta \Omega \tau$. This state is usually called
Schr\"odinger cat state, consisting of an entangled bipartite
system correlating microscopic and mesoscopic quantum states. A
measurement of the ionic internal state will project the motional
state in
\begin{eqnarray}
\label{oddevencats} \frac{1}{\sqrt{2}} \big( | \alpha \rangle \pm
| - \alpha \rangle \big) ,
\end{eqnarray}
usually called even and odd coherent states, if the outcome of the
detection was $| \downarrow \rangle$ or $| \uparrow \rangle$,
respectively. Bigger entangled and superposition states, involving
different coherent states for the case of $N$ ions, were
considered recently by the same
authors~\cite{cation,santiagoreview}.

The case $k=2$, the excitations occuring in the second red and
second blue sidebands, can offer us other interesting nonclassical
states. The operator of Eq.~(\ref{displacement}) turns into
$S(\xi)$, a squeezing operator with squeezing parameter $\xi$,
that creates two-photon coherent states\cite{Yuen}
\begin{equation}
| \xi \rangle = e^{ ( \xi a^{\dagger 2} - {\xi}^* a^2)} | 0
\rangle \, .
\end{equation}
If, again, we choose $N=1$ (for $k=2$) and an initial state $|
\downarrow \rangle | 0 \rangle = ( | + \rangle + | - \rangle ) /
\sqrt{2}$, the final state after an interaction time $\tau$ is
\begin{eqnarray}
\label{squeezedcats} \frac{1}{\sqrt{2}} \big( | + \rangle | \xi
\rangle + | - \rangle | - \xi \rangle \big) ,
\end{eqnarray}
where $| \xi \rangle$ and $| - \xi \rangle$ are squeezed vacuum
states with $\xi = i \eta^2 \Omega / 2$. These squeezing
parameters, $\xi = r e^{i \phi}$ and $- \xi = r e^{i ( \phi + \pi
) }$, will produce the same compression $e^{-r}$ in the motional
quadratures associated with the orthogonal directions $\phi / 2$
and $\phi / 2 + \pi / 2$, respectively. The state of
Eq.~(\ref{squeezedcats}) is, then, an entangled state correlating
the microscopic states $| \pm \rangle$ with the orthogonally
squeezed vacuum states $| \pm \xi \rangle$, produced with a single
bichromatic Raman laser pulse. When comparing the entangled states
of Eq.~(\ref{cats}) and those of Eq.~(\ref{squeezedcats}), note
that coherent states are considered as "classical" states and
squeezed states as "nonclassical" ones \cite{Knight}. Going to
higher values of $N$ will produce, as can be seen from
Eq.~(\ref{evoltime}), multiple entanglement of different
microscopic internal states with different squeezed vacuum states.
Note, also, that it is possible to produce squeezing in a chosen
motional quadrature without affecting or entangling the internal
degrees of freedom~\cite{nonclassicalion}. For achieving this
goal, it is enough to consider an initial state that contains one
of the electronic eigenstates, say $|j,m\rangle_x | 0 \rangle$,
that in the case of the previous example ($N = 1$) can be written
as $|\pm \rangle | 0 \rangle$.

We now consider the  simultaneous application of four Raman laser
fields on $N$ trapped ions ($N > 2$). Their frequencies are chosen
as
\begin{eqnarray}  \label{freq1}
\omega_{I} &=& \omega_0 +k \nu\,  , \quad
 \quad
\omega_{II} =\omega_0 -k\nu\, ,\nonumber \\ \omega_{III} &=&
\omega_0 +k \nu_r\, , \quad
 \quad
\omega_{IV} =\omega_0 -k\nu_r\, ,
\end{eqnarray}
where $\nu_r=\sqrt{3}\nu$ is the frequency associated with the
stretch vibrational mode. By so doing, we excite simultaneously
the $k$-th red and the $k$-th blue vibronic sidebands of the CM
and the stretch mode. The amplitudes of the fields exciting the CM
mode are taken as equal and the ones exciting the stretch mode
follow a similar condition. We take the phases of all Raman laser
pulses equal to $\varphi=k\pi/2$. The effective Hamiltonian, after
following the same steps as in the case of the bichromatic
illumination, reads
\begin{eqnarray}
\label{tetraHamiltonian} H_{\rm int} = && \frac{2 \hbar }{k!} \,\,
{J}_x \lbrack \eta^k \Omega (a^k + a^{\dagger k } ) + \eta_r^k
\Omega_r ( b^k + b^{\dagger k }) \rbrack ,
\end{eqnarray}
where $b$ and $b^\dagger$ are the annihilation and creation
operator of the stretch mode and $\eta_r = \eta / \sqrt[4]{3}$ its
associated Lamb-Dicke parameter. $\Omega$ and $\Omega_r$ are the
different coupling constants associated with the CM and stretch
mode excitations, respectively. Using a similar reasoning, it is
easy to see that the time evolution operator is now given by
\begin{equation}\label{evol1}
U(t) = \sum_{j,m } D_k ( m \Omega_k t ) D_{kr} ( m \Omega_{kr} t )
| {j,m} \rangle_{x \,\,x} \langle {j,m} | .
\end{equation}
Here, $\Omega_{kr} = 2 i \eta_r^k \Omega_r /k!$ and
\begin{equation}
D_{kr}(\chi_r) = e^{\chi_r b^{\dagger k} -\chi_r^* b^k} .
\end{equation}
In what follows we will restrict us to the case $k=1$. We can say,
then, that Eq.~(\ref{evol1}) shows the possibility of producing
simultaneous displacements in the CM and stretch mode conditioned
to the collective internal state.

Let us consider the particular case $N=2$ and an initial state $|
\downarrow \downarrow \rangle | 0 0 \rangle = \frac{1}{2}( |
\phi_1 \rangle + | \phi_2 \rangle + | \phi_3 \rangle + | \phi_4
\rangle ) | 0 0 \rangle $, where
\begin{eqnarray}
| \phi_{1,2} \rangle = \frac{1}{2} ( | \downarrow \downarrow
\rangle \pm | \downarrow \uparrow \rangle \pm | \uparrow
\downarrow \rangle + | \uparrow \uparrow \rangle ) \nonumber
\\ | \phi_{3,4} \rangle = \frac{1}{2} ( | \downarrow \downarrow
\rangle \pm | \downarrow \uparrow \rangle \mp | \uparrow
\downarrow \rangle - | \uparrow \uparrow \rangle )
\end{eqnarray}
are the eigenvectors of $J_x = \sum_{j=1}^2 (
{\sigma_{j}}^{\dagger} + \sigma_{j} ) / 2$ (spin 1) with
eigenvalues $\lambda_{1,2} = \pm 1$ and $\lambda_{3,4} = 0$,
respectively. The application of the four Raman laser fields
during a time $\tau$ will produce the entangled state
\begin{eqnarray}
\label{gatogrande} \frac{1}{2} ( | \phi_1 \rangle |\alpha , \beta
\rangle + | \phi_2 \rangle | - \alpha , - \beta \rangle + | \phi_3
\rangle
| 00 \rangle + | \phi_4 \rangle | 00 \rangle ) && , \nonumber \\
&&
\end{eqnarray}
with $\alpha = 2 i \eta \Omega \tau$ and $\beta = 2 i \eta_r
\Omega_r \tau$. This is a more sophisticated Schr\"odinger cat
state, correlating several microscopic states with the vacuum and
different mesoscopic states in the CM and stretch motional modes.
Measuring the internal state $| \downarrow \uparrow \rangle$ or $|
\uparrow \downarrow \rangle$ will project the motional state into
\begin{equation}
\label{entcohstates} \frac{| \alpha , \beta \rangle - | - \alpha ,
- \beta \rangle} {\sqrt{2 ( 1 - e^{- 2 | \alpha |^2 - 2 | \beta
|^2})}} .
\end{equation}
These states, called entangled coherent states, own special
entanglement properties~\cite{hirota}, and have found diverse
applications in the domain of quantum information
theory~\cite{vanEnk,munro,kim}. In the language of the {\it
Gedankenexperiment} after Schr\"odinger~\cite{Schrodinger}, the
state of Eq.~(\ref{entcohstates}) represents the simultaneous
occurrence of two alife cats and two dead cats. To our knowledge
it is the first proposal for producing such entangled states in
normal modes of $N$ trapped ions, with homogeneous illumination,
involving quantum mechanical correlations between two classical
states~\cite{Knight2}. It is clear that, by imposing $\eta \Omega
= \eta_r \Omega_r$, we could have made $\alpha = \beta$. If in
Eq.~(\ref{gatogrande}) we measure $| \downarrow \downarrow
\rangle$ or $| \uparrow \uparrow \rangle$, we will project the
motional state into
\begin{equation}\label{sleepycat}
{\cal N} \left ( | \alpha , \beta \rangle + | - \alpha , -\beta
\rangle \pm 2 | 00 \rangle \right) \, ,
\end{equation}
respectively, where ${\cal N} $ is a normalization constant. A
more elaborated way of producing entangled coherent states is to
consider the initial state $\frac{1}{\sqrt{2}} (| \downarrow
\downarrow \rangle + | \uparrow \uparrow \rangle) =
\frac{1}{\sqrt{2}} (| \phi_1 \rangle + | \phi_2
\rangle)$~\cite{sorensenBell,SMZBell}. The application of the four
Raman laser fields during a time $\tau$ generates the state
\begin{eqnarray}
\label{bigcats} \frac{1}{\sqrt{2}} \big( | \phi_1 \rangle | \alpha
, \beta \rangle + | \phi_2 \rangle | - \alpha, - \beta \rangle
\big) .
\end{eqnarray}
The subsequent detection of the state $| \downarrow \downarrow
\rangle$ will produce the state
\begin{equation}
\label{entcohstates+} \frac{| \alpha , \beta \rangle + | - \alpha
, - \beta \rangle} {\sqrt{2 ( 1 + e^{- 2 | \alpha |^2 - 2 | \beta
|^2})}} ,
\end{equation}
and the measurement of the state $| \uparrow \uparrow \rangle$
will reproduce the entangled coherent state of
Eq.~(\ref{entcohstates}).

It is easy to generalize this procedure to $n$ vibrational normal
modes in an $N$ ion system, by using different bichromatic
excitation fields with frequencies $\omega_o \pm k \nu_{n}$,
$\nu_n$ being the frequency of the $n^{\rm th}$ normal mode. By so
doing, we could obtain more sophisticated entangled mesoscopic
superpositions.

In conclusion, we have shown that resonant bichromatic and
tetrachromatic excitations of $N$ trapped ions can be an important
tool for producing different families of nonclassical states,
involving the ionic internal and external degrees of freedom.
Generating bigger nonclassical states, and studying their decay
properties, should help us to understand better the border between
the quantum and the classical world. At the same time, interesting
applications may appear, specially the ones related to the field
of quantum information. In particular, for the case of a single
ion and a bichromatic illumination, we have shown how to produce
Schr\"odinger cat states and entangled squeezed states by means of
a single bichromatic Raman laser pulse. In the case of two ions
and tetrachromatic excitation, we have shown how to produce, in a
straightforward manner, entangled coherent states in two motional
normal modes, and other states involving multiple entanglement.

This work was partially supported by the Conselho Nacional de
Desenvolvimento Cient{\'\i}fico e Tecnol\'ogico (CNPq), the
Funda\c{c}\~ao de Amparo a Pesquisa do Estado do Rio de Janeiro
(FAPERJ),  the Programa de Apoio a N\'ucleos de Excel\^encia
(PRONEX) and Funda\c{c}\~ao Jos\'e Bonif\'acio (FUJB).


\begin{references}

\bibitem{ioncats} C. Monroe, D. M. Meekhof, B. E. King, and
D. J. Wineland, Science {\bf 272}, 1131 (1996).

\bibitem{CQEDcats} M. Brune, E. Hagley, J. Dreye, X. Ma\^{\i}tre,
A. Maali, C. Wunderlich, J. M. Raimond, and S. Haroche, Phys. Rev.
Lett. {\bf 77}, 4887 (1996).

\bibitem{fourionent}  C. A. Sackett, D. Kielpinski, B. E. King,
C. Langer, V. Meyer, C. J. Myatt, M. Rowe, Q. A. Turchette, W. M.
Itano, D. J. Wineland and C. Monroe, Nature {\bf 404}, 256 (2000).

\bibitem{threeatoment} A. Rauschenbeutel, G. Nogues, S. Osnaghi,
P. Bertet, M. Brune, J. M. Raimond, and S. Haroche, Science {\bf
288}, 2024 (2000).

\bibitem{reviewions} D. J. Wineland, C. Monroe, W. M. Itano,
D. Leibfried, B. E. King, D. M. Meekhof, J. Res. NIST {\bf 103},
259-328 (1998).

\bibitem{reviewCQED} J. M. Raimond, M. Brune, and S. Haroche, Rev.
Mod. Phys. {\bf 73}, 565 (2001).

\bibitem{Zurek}  See, for example, J. P. Paz and W. H. Zurek in
{\it Coherent Matter Waves}, Proceedings of Les Houches Summer
School, Session LXXII, edited by R. Kaiser, C. Westbrook, and F.
David (EDP Sciences, Les Ulis; Springer-Verlag, Berlin, 2001).

\bibitem{Schrodinger}  E. Schr\"{o}dinger, Naturwissenschaften {\bf 23},
807(1935).

\bibitem{decoherenceions} C. J. Myatt, B. E. King, Q. A. Turchette,
C. A. Sackett, D. Kielpinski, W. M. Itano, C. Monroe, and D. J.
Wineland, Nature {\bf 403}, 269 (2000).

\bibitem{cation} E. Solano, R. L. de Matos Filho, and N. Zagury,
Phys. Rev. Lett. {\bf 87}, 060402 (2001).

\bibitem{santiagoreview} E. Solano, R. L. de Matos Filho, and N.
Zagury, Lecture Notes in Physics, 2001, no. {\bf 575}, pp. 14-28,
Springer-Verlag.

\bibitem{king}  B. E. King, C.~S. Wood, C.~J. Myatt, Q.~A. Turchette,
D.~Leibfried, W.~M. Itano, C.~Monroe, and D.~J. Wineland, Phys.
Rev. Lett. {\bf 81}, 1525 (1998).

\bibitem{James} D.~F.~V.~James, Appl. Phys. B {\bf 66}, 181 (1998).

\bibitem{sashavogel} S. Wallentowitz and W. Vogel, Phys. Rev. A
{\bf 54}, 3322 (1996).

\bibitem{moyacessasashavogel} H. Moya-Cessa, S. Wallentowitz, and
W. Vogel, Phys. Rev. A {\bf 59}, 2920 (1999).

\bibitem{vogelruynet}  W. Vogel and R.L. de Matos Filho, Phys. Rev. A{\bf 52},
4214 (1995).

\bibitem{Yuen} H. P. Yuen, Phys. Rev. A{\bf 13}, 2226 (1971).

\bibitem{Knight} S.-C. Gou, J. Steinbach, and P. L. Knight, Phys. Rev. A
{\bf 55}, 3719 (1997).

\bibitem{nonclassicalion} D. M. Meekhof, C. Monroe, B. E. King, W. M. Itano, and D. J.
Wineland, Phys. Rev. Lett. {\bf 76}, 1796 (1996).

\bibitem{hirota} O. Hirota and M. Sasaki, LANL e-print
quant-ph/0101018.

\bibitem{vanEnk} S. J. van Enk and O. Hirota, Phys. Rev. A {\bf
64}, 022313 (2001).

\bibitem{munro} W. J. Munro, G. J. Milburn, and B. C. Sanders,
Phys. Rev. A {\bf 62}, 052108 (2000).

\bibitem{kim} H. Jeong and M. S. Kim, LANL e-print
quant-ph/0111015.

\bibitem{Knight2} J. Steinbach, J. Twamley, and P. L. Knight,
Phys. Rev. A {\bf 56}, 4815 (1997).

\bibitem{sorensenBell} A. S{\o}rensen and K. M{\o}lmer, Phys. Rev. Lett.
{\bf 82}, 1971 (1999).

\bibitem{SMZBell}  E. Solano, R. L. de Matos Filho, and N. Zagury,
Phys. Rev. A, {\bf 59}, R2539 (1999);  {\bf 61}, 029903(E) (2000).

\end{references}
\end{document}